\documentclass[reprint,amsmath,amssymb,aps,twocolumn,prb]{revtex4-1}
\usepackage{xcolor}
\usepackage{graphicx}
\usepackage{dcolumn}
\usepackage{bm}

\begin{document}

\preprint{APS/123-QED}

\title{Periodic  arrays of intercalated atoms in 
twisted bilayer graphene: \\ an {\it ab initio} investigation}

\author{R. H. Miwa}
\affiliation{Instituto de F\'{\i}sica, Universidade Federal de   Uberl\^andia,
C.P. 593,   38400-902, Uberl\^andia, MG, Brazil.}
\email{hiroki@infis.ufu.br}
\author{P. Venezuela}
\affiliation{Instituto de F\'{\i}sica, Campus da Praia Vermelha, Universidade Federal Fluminense, Niter\'oi, RJ, Brazil}
\email{vene@if.uff.br}
\author{Eric Su\'arez Morell}
\affiliation{Departamento de F\'{i}sica, Universidad T\'{e}cnica
Federico Santa Mar\'{i}a, Casilla 110-V, Valpara\'{i}so, Chile}
\email{eric.suarez@usm.cl}

\begin{abstract}

We have performed an {\it ab initio} investigation  of transition metals (TMs = 
Mo, Ru, Co, and Pt) embedded in twisted bilayer graphene (tBG) layers. Our 
total energy results reveal that,  triggered by the misalignment between the 
graphene layers, Mo and Ru atoms may form a quasi-periodic  
(triangular) array of intercalated atoms. In contrast, the formation of  those 
structures is not expected for the other TMs, Co and Pt  atoms. 
The net magnetic moment (m) of Mo and Ru atoms may be 
quenched upon 
intercalation,  depending on the stacking region (AA or AB).
For instance, we find a magnetic moment of 0.3\,$\mu_{\rm B}$ (1.8\,$\mu_{\rm 
B}$) for Ru atoms intercalated between the AA (AB) regions of the stacked twisted layers. Through 
simulated scanning tunneling microscopy (STM) images,  we verify that the 
presence of intercalated TMs can be identified by the formation of bright 
(hexagonal) spots lying on the graphene surface. 

\end{abstract}

\maketitle

\section{Introduction}

The synthesis of graphene and the experimental measurements of its outstanding properties\,\cite{novoselovNat2005} opened a pathway to explore and synthesize other single layer or few layers two dimensional (2D) materials such as hexagonal boron nitride (hBN), molybdenum disulphide\,\cite{Wang_2012,Heinz_2010} and phosphorene, a single layer of black phosphorus\,\cite{Castellanos_2014}. Furthermore the concept of van der Waals (vdW) heterostructure has emerged\,\cite{geimNature2013}, layers of 2D  materials, which interact by means of vdW forces, 
are piled up, 
aiming to provide a suitable set of electronic and structural properties. Recently heterostructures conformed by vertically stacked layers of graphene/hBN/graphene  have been successfully synthesized\,\cite{mishchenkoNatNanotech2014}, in these cases the different lattice constants and/or a relative rotation angle (RRA) between layer lead to moir\'e patterns that are observed in scanning tunneling microscopy (STM) images.

The synthesis of few layer graphene by chemical methods\cite{Varchon_2008,Reina2009}, for instance, shows quite often a rotation between graphene layers with respect to the Bernal or AB stacking. This gives rise to a moir\'e pattern with a periodicity that can be associated with the RRA between layers. This structure is known as twisted bilayer graphene (tBG) and it has been extensively studied due to its peculiar electronic properties\cite{Lopes_2007,varchonPRB2008,duNature2009,Trambly_2010,Shallcross_2010,Morell_2010,Havener_2012,Lu_2013,Morell_2013,Shallcross_2013,Zhang_2014,Chung_2015}.

 In tBG, there is a so-called AA region
where the stacking is similar to AA bilayer graphene, with one atom
exactly above another atom of the other layer, and another region with AB stacking, the AB region. Between the AA and AB regions the stacking looks like two random slipped layers.
The size of the tBG unit cell (UC) and of the AA region is increased for low RRA and the system can be thought as a large AA region surrounded by
AB areas, as depicted in Fig.\,\ref{xyz}(a). Low energy electrons are localized
in the AA region as they can not easily penetrate the barrier formed
by the AB stacking\cite{Luican_2011,Trambly_2012a,Lopes_2012}. The local density
of states is significantly larger at the AA site for very low angles and this is precisely the region one
observe in STM images\cite{varchonPRB2008,millerPRB2010,othaPRL2012}. The electronic spectrum also depends on the RRA. 
For large  RRA ($\thicksim 20^{\rm o}$), the system behaves as if the two layers were uncoupled with a linear energy dispersion like in monolayer graphene. Below $20^{\rm o}$
   there is a constant reduction of the Fermi velocity of charge carriers and close to $1^{\rm o}$ 
 the spectrum shows zero-velocity flat bands. Scanning tunneling spectroscopy (STS) measurements have shown angle dependent Van Hove singularities(vHs) near the fermi level\cite{duNature2009} consistent with the behavior predicted by theoretical methods\cite{Lopes_2007,varchonPRB2008,Trambly_2010,Shallcross_2010,Morell_2010}.

There have been recently several proposals to exploit the properties of graphene by intercalation of metals. An enhancement of the out of plane magnetic anisotropy have been observed in intercalated Cobalt atoms between graphene and Ir(111) substrate\,\cite{deckerPRB2013,vlaicAPL2014}, the anisotropy also depends on the specific site of the unit cell where the Co atoms are deposited\,\cite{deckerPRB2013}. In some cases the intercalated Co atoms forms a quasi-periodic array of trimers between graphene adlayer and the SiC(0001) surface\,\cite{limaChemMat2014}. Besides, the geometry of adatoms adsorbed on the graphene surface, might play a very important role on the formation of quantum spin and anomalous Hall phases in graphene\,\cite{acostaPRB2014}.


In this work we propose to take advantage of the moir\'e structure in twisted 
bilayer graphene to create a periodic array of intercalated 
transition metal(TM) atoms. If there is any energetic preference for the atoms 
to be localized in a given region between the layers of the twisted unit cell it 
might be possible to control the distance between them by manipulating the RRA 
between layers,  allowing or not the interaction between atoms and controlling 
the electronic/magnetic properties of the system. 

We examine, based on {\it ab initio} total
energy calculations, the energetic stability and the electronic
properties of TMs (TM=Mo, Ru, Co, and Pt) intercalated between tBG
layers. We found that Mo and Ru atoms exhibit an energetic preference for the AA
stacking region (TM$_{\rm AA}$) of the twisted bilayer unit cell, while no such a preference is found
for Co and Pt. This suggest that Mo and Ru intercalated atoms might form a 
 quasi-periodic triangular array
embedded in the twisted layers. At the same time the net magnetic moment  of
Mo, Ru and Co atoms are somewhat quenched  upon their intercalation, for
instance,  we found a magnetic moment of 0.3\,$\mu_{\rm B}$ for Ru adatoms intercalated in the AA
stacking region. Through STM simulation, we show that there is an increase of the electronic density of states on the (graphene) C atoms bonded to the TM incorporation sites.

\section{Method}

\subsection{Geometry of the Unit cell.}
\begin{figure}[thpb]
   \centering
\includegraphics[width= 7.5cm]{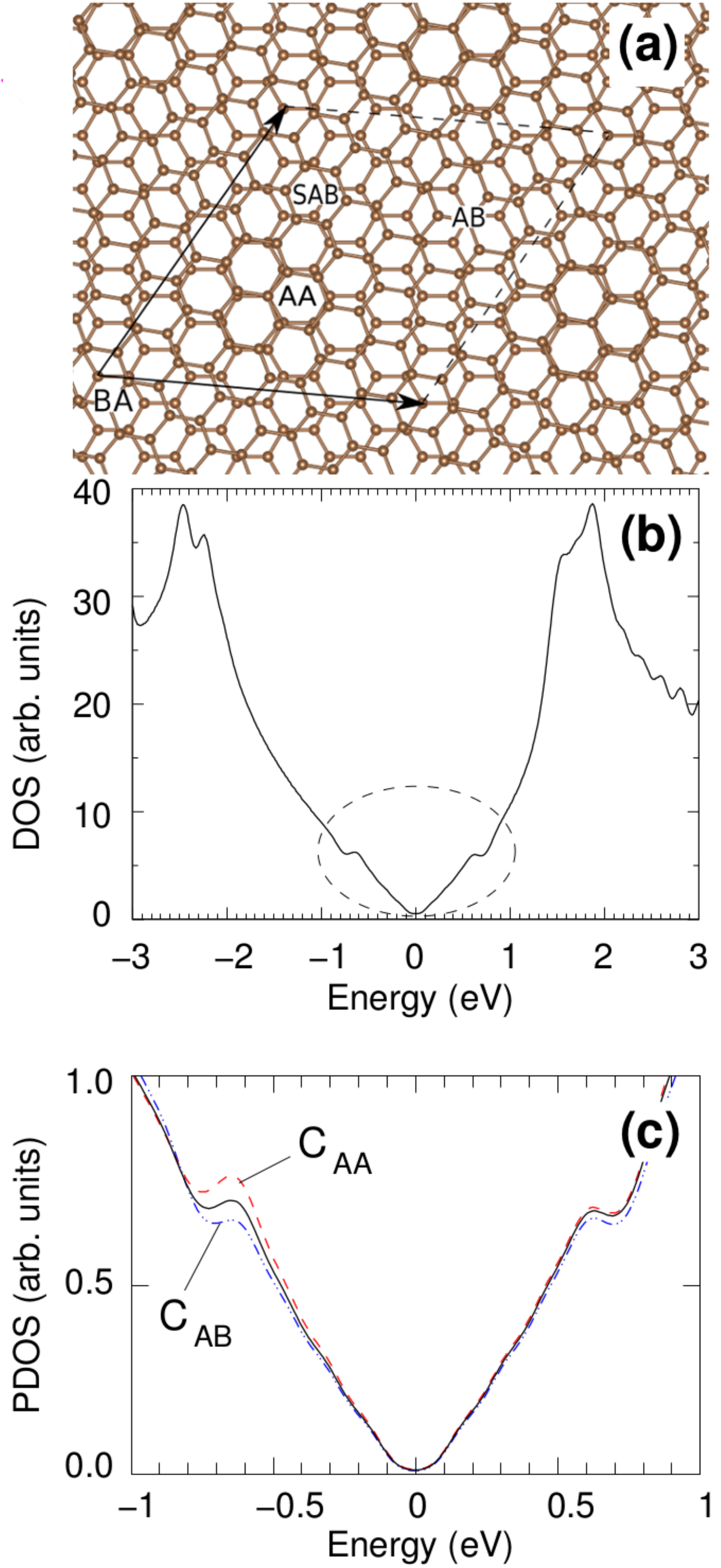}
\caption{\label{xyz}  (a) Structural model of twisted graphene bilayer with RRA
  of 9.34$^\circ$ (tGB-148). AA, AB, BA and SAB stacking regions are indicated (see text) and the arrows represent the supercell lattice vectors. 
(b) tBG-148 total electronic density of states (DOS) and (c) projected DOS
  (PDOS) on the carbon atoms near the AB (C$_{\rm AB}$) and AA
  (C$_{\rm AA}$) stacking regions.  In (b) and (c), we have 
considered a gaussian broadening of 0.1\,eV in the discrete electronic energies, and the Fermi levels were set to 
zero.}
\end{figure}

The system we studied is composed of two coupled graphene layers with
a given RRA between them. We built a commensurate unit cell following a
procedure exposed in several previous
works\cite{Campanera_2007,Morell_2011b}. Starting from a stacked AB
bilayer graphene one rotate a lattice point to an equivalent location,
a vector $\vec{r}=m \vec{a}_{1} + n \vec{a}_{2} $ is rotated to
$\vec{t}=n \vec{a}_{1} + m \vec{a}_{2}$, where $\vec{a}_{1}$ and
$\vec{a}_{2}$ are the graphene bilayer lattice vectors;
\textit{n},\textit{m} are integers. These integers are the only
information needed to completely characterize the unit cell, the
rotating angle, the unit cell vectors, and the numbers of atoms in the
UC can be expressed in terms of them\cite{Morell_2010,Lopes_2012}.
The UC constructed this way contains one site, known as AA, where one
atom is exactly above another atom of the other layer, there is another site, AB,
where an atom of the top layer is exactly in the center of a lower layer
hexagon, we have also a BA site, equivalent to the AB site but with the difference that an atom in the lower layer is in the center of an hexagon of the upper layer, there is also another area of the UC where the stacking is none of the above mentioned we call it slipped AB (SAB), see Fig. \ref{xyz}(a).  As the UC gets larger,
very small RRA, the number of atoms with almost exact AA or AB character
increases\cite{Campanera_2007} creating large regions with almost the same
stacking. The distance between two AA regions (periodicity of the superlattice) can be calculated by the equation $D= a_{0} / 2 \times \sin(\theta/2)$, where $a_{0}=2.46$ \AA\, and $\theta$ is the RRA between layers.

\subsection{DFT calculations}
The calculations were performed within the density functional theory (DFT) 
approach, as implemented in the Quantum-Espresso package~\cite{espresso}. The 
Kohn-Sham orbitals were expanded in a plane-wave basis set, with an energy 
cutoff of 35~Ry.  The twisted graphene bilayers were simulated considering 
supercells with 148 (tGB-148) and 244 (tGB-244) atoms, corresponding to RRA 
($\theta$) of $9.34^{\rm o}$ and $7.34^{\rm o}$, respectively, and vacuum 
regions of 14\,\AA. 
calculations. The atomic positions were fully relaxed, by including van der 
Waals (vdW) interactions within  the self-consistent vdW-DF approach as 
described in Refs.~\cite{dionPRL2004,thonhauserPRB2007,perezPRL2009}.  For the 
structural relaxations we have considered a force convergence tolerance of 
15\,meV/\AA \, and the Brillouin zone (BZ) was sampled by $4\times 4\times 1$ 
($2\times 2\times 1$)  {\bf k}-points in the case of tGB-148 (tGB-244). After 
the structural relaxations, spin polarized calculations were performed with 
$8\times 8\times 1$ ($5\times 5\times 1$)  {\bf k}-points BZ sampling in the 
case of tGB-148 (tGB-244).

\section{Results}

We begin examining the electronic and structural properties of the twisted 
bilayer graphene crystal in comparison with recent experimental and theoretical 
findings. In Fig.~\ref{xyz}(a) we show the structural model of tGB-148, 
indicating the AA and AB stacking regions. At equilibrium, the averaged 
interlayer distance ($\rm\Delta z$) is 3.679\,\AA; slightly larger than the 
distance obtained for perfect bilayer graphene (AB or Bernal stacking). There 
are no chemical bonds between the graphene sheets, the interlayer interaction is 
governed by vdW forces, we found a binding energy of 27\,meV/C-atom very close 
to the experimental value of graphite\cite{Benedict_1998,Liu_2012}. In the case 
of tGB-148, RRA = $9.34^{\rm o}$, a slight renormalization of the Fermi velocity 
and the presence of vHs, associated with the M point of the BZ are 
expected\cite{Morell_2010}. The energy of these vHs depends on the RRA, 
diminishing as the angle gets 
lower\,\cite{duNature2009,Luican_2011,brihuegaPRL2012}.   We found that the vHs 
lie at $\pm 0.64$\,eV with respect to the Dirac point, as shown in 
Fig.~\ref{xyz}(b) in agreement with previous experimental 
results\cite{brihuegaPRL2012}.


The moir\'e structure gives rise to a periodic potential that have a huge 
impact on the electronic dispersion and at the same time promotes a charge 
density redistribution confining low energy electrons to the AA region, 
producing the bright spot observed in STM experiments on graphite or 
tBG\,\cite{varchonPRB2008,millerPRB2010,othaPRL2012}. 

The origin of these bright spots were the subject of a debate not so long ago\cite{Pong_2005}. Two theories were vindicated:  (i) there is an  increase of the electronic density of states around the AA region, when compared with the AB area 
and (ii) the AA and AB regions present different equilibrium
geometries,  and the interlayer distances $\rm\Delta z$ (topographic
corrugation) are different. Both arguments are valid and indeed, (i) can be verified in
Fig.~\ref{xyz}(c), by projecting the electronic DOS on the C atoms near the AA
(C$_{\rm AA}$) and AB (C$_{\rm AB}$) stacking regions. The electronic density of
states is larger on C$_{\rm AA}$.  In addition, comparing the
equilibrium geometries,  we find that the interlayer distance in the AA region ($\rm\Delta
z_{AA} = 3.692$\,\AA) is larger by 0.04\,\AA\ than in the AB one
 ($\rm\Delta z_{AB} =
3.653$\,\AA), as proposed in (ii). By reducing the twist-angle to 7.34$^\circ$
(tGB-244), the lateral distance between the AA and AB regions increases, when
compared with tGB-148, and  we find practically the same (averaged) interlayer
distance, 3.681\,\AA. However, the  interlayer distance in AA region increases to
$\rm\Delta z_{AA} = 3.765$\,\AA, while in the AB region we find practically the same distance
($\rm\Delta z_{AB} = 3.650$\,\AA). These small changes in the interlayer distances suggest that the origin of the STM images is mainly electronic rather than geometrical\cite{Cisternas_2012}.

%

We found pertinent to investigate whether this behavior facilitates the creation of periodic arrays of  intercalated
atoms between twisted graphene layers. To such end we studied the energetic stability of transition metal atoms (TMs: Mo, Ru, Co, and Pt) intercalated at different sites of the tBG unit cell. A measure of the energetic stability is the binding energy of the TMs embedded in the twisted systems, the binding energy ($E^b$) can be obtained from:
$$
E^b =  E[{\rm tBG}] + E[{\rm TM}] - E[{\rm TM/tBG}].
$$
Where $E[{\rm TM/tBG}]$ is the total energy of the final system, a tBG
intercalated by a TM atom. $E[{\rm tBG}]$  and
$E[{\rm TM}]$ are the total energies of the separated components, 
a tBG and an isolated TM atom, respectively.


We calculated the binding energy at several locations within the UC and found 
that for Co and Pt the differences are small, there is not any clear preference 
for any region. However for Mo and Ru the difference can be as large as 1 eV, 
and the highest binding energy is obtained in the AA region and the lowest in 
the AB one. In the rest of the UC the results are in between these two extreme 
values.  

In addition we have calculated the binding energy but with the TM atoms adsorbed 
on the surface site(an atom on top of the upper layer). We obtained, for the 
tBG-148 structure, $E^b$ of 3.48, and 2.47 for Mo and Ru adatoms lying on the 
hollow site of the AA region, and 1.44\,eV for Pt adatom on the bridge site. 
These values are significantly smaller than the ones obtained for the 
intercalated atoms.

\begin{table}[h]
\caption{\label{energy} Binding energies ($E^b$ in eV/TM-atom) of TMs intercalated
in the AA and AB stacking regions of twisted graphene bilayer, tBG-148 and
tBG-244.}
\begin{ruledtabular}
\begin{tabular}{lcccc} 
\multicolumn{1}{c}{ } &
\multicolumn{2}{c}{$\theta = 9.34^\circ$ (tBG-148)} &
\multicolumn{2}{c}{$\theta = 7.34^\circ$ (tBG-244)} \\
 TM          &   AA    &   AB   &  AA   &  AB  \\
\hline
  Mo        &  5.79   &  4.64  & 5.88  & 4.73 \\
  Ru        &  3.76   &  3.16  & 3.80  & 3.20 \\
  Co        &  2.74   &  2.73  & 2.74  & 2.73 \\
  Pt        &  1.66   &  1.63  & 1.77  & 1.71 \\
\end{tabular}
\end{ruledtabular}
\end{table}
Table~I shows the values of the binding energy for the AA and AB regions for all the intercalated TM considered. There is an energetic preference (higher binding energies) for Ru and Mo atoms intercalated in the  AA region. For the tBG-148 system, the Mo$_{\rm AA}$(Ru$_{\rm 
AA}$) structure is more stable than Mo$_{\rm AB}$(Ru$_{\rm 
AB}$) by 1.15\,eV (0.60\,eV). Such energetic preference has been 
maintained in tBG-244. That is, the binding energies of Mo$_{\rm AA}$ 
and Ru$_{\rm AA}$ are higher by 1.15 and 0.60\,eV, 
respectively, compared with the ones of Mo$_{\rm AB}$ and Ru$_{\rm 
AB}$. On the other hand,  the  energetic preference for the AA stacking 
region reduces to 0.03\,eV (0.06\,eV) for Pt$_{\rm AA}$/tBG-148 (Pt$_{\rm 
AA}$/tBG-244), while the Co$_{\rm AA}$ and Co$_{\rm AB}$ systems 
present practically the same binding energy. In the appendix we show the results of the binding energies for Ru and Mo in other regions of the UC. In general within the AA region the differences are small (around 1\%) and in the SAB and AB regions the binding energies are around 5\%  and 20 \% lower, respectively, than in the AA region.

\begin{figure}[thpb]
\includegraphics[width= 8.5cm]{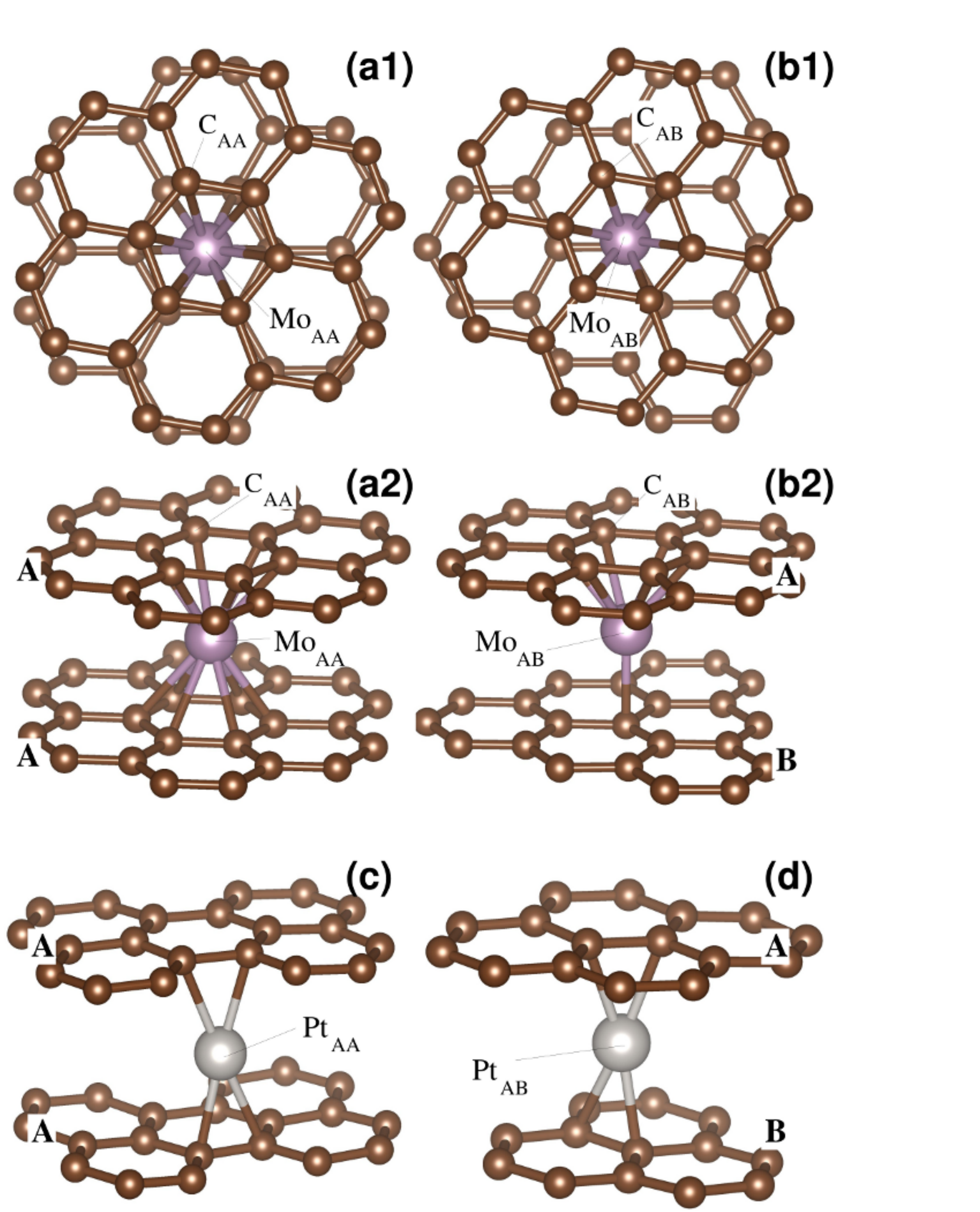}
\caption{\label{models}  Equilibrium geometry of  Mo$_{\rm 
AA}$/tBG, top-view (a1) and side-view (a2),  and Mo$_{\rm AB}$/tBG TM(AB), 
top-view (b1) and side-view (b2). The same structural model has been obtained 
for 
Ru/tBG and Co/tBG.  Equilibrium geometry of Pt$_{\rm AA}$/tBG (c) and Pt$_{\rm 
AB}$/tBG (d).}
\end{figure}

In Figs.~\ref{models}(a) and \ref{models}(b) we show the structural models of 
Mo atoms intercalated in the regions AA and AB, the structural models of Ru and 
Co are very similar. In the AA stacking region the TM 
atom lies on the center of one of the empty hexagons (Fig.~\ref{models}(a)), in this 
case the TM atom is bonded to 12 carbon atoms\cite{coordination}. On the other 
hand, in the AB stacking region (Fig.~\ref{models}(b)), the TM atom lies on top 
of one atom of one layer, and on the hollow site of the other one. In this case, 
the coordination of the TM atom is reduced to seven. The 
interlayer distance, upon the presence of intercalated Mo and Ru atoms, 
$\rm\Delta z = 3.73$\,\AA, is slightly larger compared with the one obtained for 
the pristine twisted system. Whereas, we find $\rm\Delta z$ around 3.85\,\AA\ 
for Mo$_{\rm AB}$ and Ru$_{\rm AB}$, thus suggesting that those 
structures induce a larger local strain when compared with their Mo$_{\rm 
AA}$ and Ru$_{\rm AA}$ counterparts.  Differently from the other TM 
atoms, the Pt atoms lie on the bridge site of the graphene layer in both 
stacking regions, Figs.\ref{models}(c) and \ref{models}(d). Thus in both cases 
Pt atoms are bonded to four carbon atoms.
 Mo and Ru atoms, as d-orbital metals, tend to have as many bonds as possible when they are in contact with organic materials while Pt atoms behave more like noble elements.

For Co atoms the different coordination in the AA and AB stacking 
region do not lead to large differences in the total energies. This is due to 
the fact that Co covalent radius is smaller than Mo and Ru radius, thus 
it is expected because the covalent bonds between Co and carbon atoms are not 
as strong as Mo(Ru)-C bonds. 

We proceed now to quantify the local strain in the twisted graphene bilayer, upon the
intercalation of the TM atoms, we calculate the deformation energy
($E^{def}$), 
$$ 
E^{def} = E^{\rm strain} - E^{\rm prist},
$$ 
here  $E^{prist}$
and $E^{\rm strain}$ represent the total energies of pristine (full relaxed) and
strained tBG-148 system, respectively. The latter term  was obtained by
considering  the atomic positions of the deformed (due to the presence of the
intercalated TM) tBG-148 structure.
The deformation energies for Mo
and Ru atoms intercalated in the AA stacking region  Mo$_{\rm AA}$  and
Ru$_{\rm AA}$ are 0.12 and 0.09\,eV respectively, while it 
increases
by about 0.1\,eV ($E^{def}$= 0.24 (0.19) \,eV) for Mo$_{\rm AB}$  (Ru$_{\rm AB}$). That is, the local strain
also contributes to the energetic preference for the AA stacking regions for these two transition metals. 
 On the other hand, for intercalated Co 
and Pt atoms, the differences are smaller, 0.024 and 0.015\,eV, 
respectively.

Summarizing the total energy results; we have found that the most stable place for Mo and Ru atoms, 
intercalated in tBG, is in the AA stacking region. This total energy difference may be 
explained by the different chemical coordination and also by the strain energy 
differences. 

Therefore our results indicate that Mo and Ru atoms may form a  
quasi-periodic triangular array when they are intercalated in twisted graphene 
layers, the most stable position is within the AA region and the energy 
differences with other regions are larger than room temperature thermal energy. 
The periodicity of these arrays can be tuned by the twist angle. 

The relative charge concentration around the AA region increases as the angle 
gets lower we expect then that the binding energy at the AA region will be 
larger when we compare with the rest of the UC for lower angles as the charge 
density should be proportional to the strength of the covalent bond between TM 
and Carbon.

\begin{figure}[thpb]
\includegraphics[width= 8.5cm]{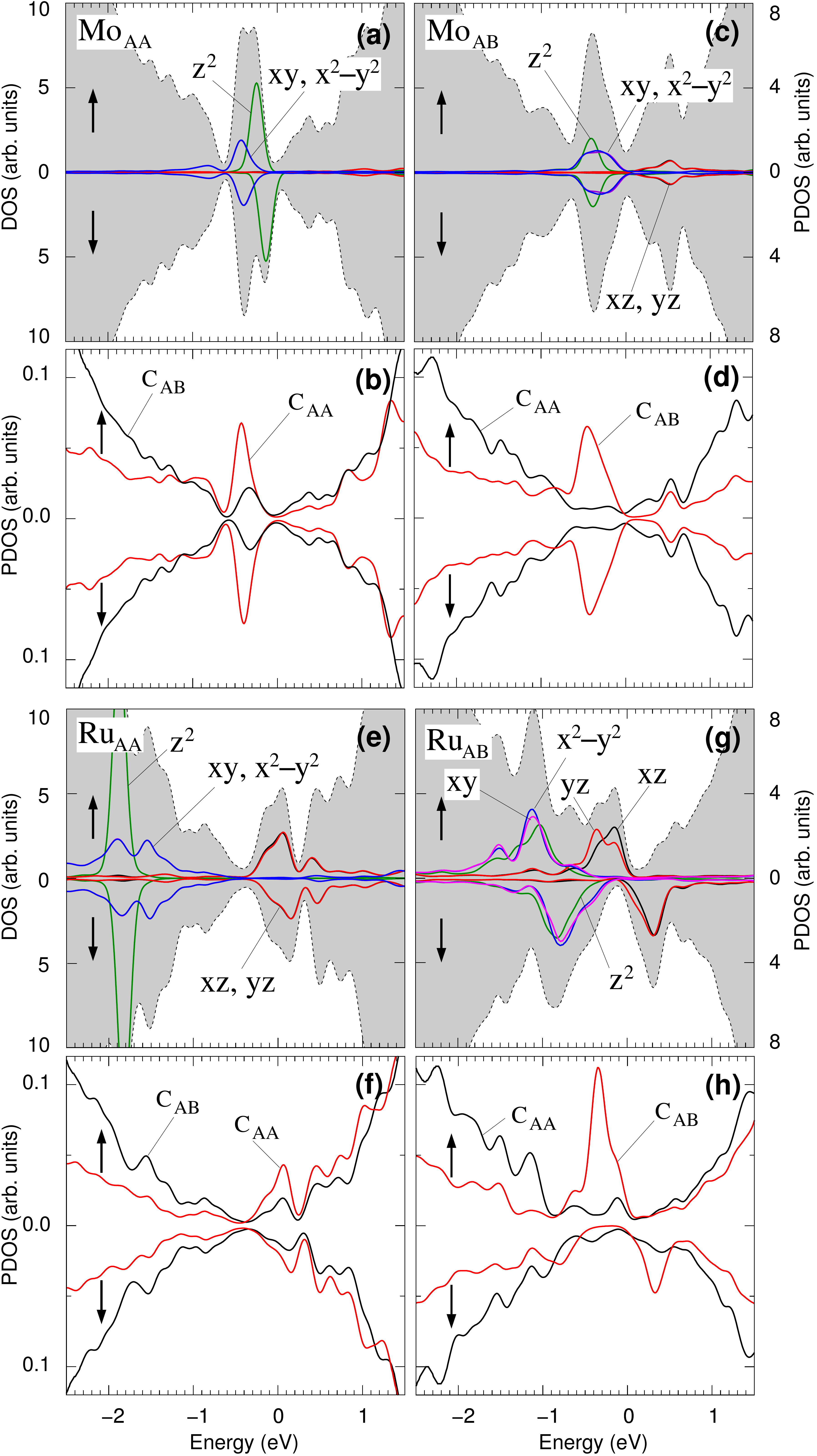}
\caption{\label{pdos}{ (left) Total density of 
states of states (DOS), shaded plots, and the projected density of states (PDOS), solid lines, of the TM-4d 
($\rm z^2, xz, yz, xy, and~x^2-y^2$) orbital, (right) projected density of 
states of C-2p$_{\rm z}$ orbitals of  C atoms of the graphene layer A (see 
Fig.\,\ref{models}) bonded to the TM (red solid line) and far from the TM (black 
solid line); for Mo/tBG-148 (a)--(d), and Ru/tBG-148 (e)--(h). The (spin 
polarized) DOS and PDOS are in arbitrary units, and the Fermi level was set to 
zero.}}
\end{figure}

We will focus now on the electronic and magnetic properties of  intercalated TM 
atoms. Mo and Ru have different magnetic moments (m) when we compare with the 
adsorbed ones, as can be noticed in Table~II. The 
magnetic moment of Mo is quenched when it is intercalated in both AA and AB stacking 
regions. Ru magnetic moment is also quenched in the more energetically stable AA 
stacking region but not in the AB region. Those changes on the 
magnetic moment are ruled by the different TM hybridization with the tBG host.
The magnetic moments of Co is almost unchanged when comparing adsorbed 
and intercalated results. This is another evidence for the fact, mentioned above, 
that Co-carbon bonds are weaker in comparison with Mo-carbon and Ru-carbon 
bonds. 


\begin{table}[h]
\caption{\label{magnetic} Magnetic moment (m in $\mu_{\rm B}$) of the TMs adsorbed
(m$_{\rm ads}$),  and intercalated in the AA and AB sites (m$_{\rm AA}$ and
m$_{\rm AB}$) of the twisted (tBG-148) graphene layers.}
\begin{ruledtabular}
\begin{tabular}{lccc} 
 TM          & m$_{\rm ads}$  &  m$_{\rm AA}$ & m$_{\rm AB}$  \\
\hline
  Mo        & 2.0    & 0.1   &  0.0  \\       
  Ru        & 1.9    & 0.3   &  1.8  \\
  Co        & 1.2    & 1.4   &  1.2 \\
  Pt        & 0.0    & 0.0   &  0.0 \\
\end{tabular}
\end{ruledtabular}
\end{table}

There is an electronic charge transfer from the Mo and Ru atoms in the AA region 
to the graphene layers. This lead to n-type doped tBG, with charge transfers of 
3.2 and $5.1\times 10^{13} {\rm cm}^{-2}$, for Mo and Ru atoms, respectively. 
Also, the local hybridization of TMs in AA and AB regions are different, as 
shown by the calculated spin-polarized electronic density of states (DOS), 
shaded plots in Fig.\,\ref{pdos}, and projected DOS (PDOS, solid lines in 
Fig.~\ref{pdos}). The Mo-4d electronic states are mostly localized below the 
Fermi level ($E_F-1$\,eV), and the net magnetic moment of 0.1\,$\mu_{\rm B}$ 
comes from a small occupation unbalance of Mo-4d$_{\rm z^2}$ orbitals, indicated 
by green solid line in Fig.\,\ref{pdos}(a).  It is worth noting that although 
the $D_{6h}$ symmetry of the intercalated Mo$_{\rm AA}$ atom is slightly missed, 
due to the twist angle, the double degeneracy of 4d$_{\rm xy}$/4d$_{\rm 
x^2-y^2}$ and  4d$_{\rm xz}$/4d$_{\rm yz}$ (levels $e$ and $e^\prime$) have been 
preserved. In addition, as presented in Fig.\,\ref{pdos}(b), there is  an 
increase of the DOS projected on the C atoms bonded to Mo$_{\rm AA}$, C$_{\rm 
AA}$ in Fig.\,\ref{models}(a2).

The  hybridization of Mo atoms in the AB region with the 
graphene layers has been strengthened due to the formation of Mo-C bond with a 
C atom of the graphene layer B [Fig.\,\ref{models}(b2)]. In this case, the 
occupation of Mo-4d$_{\rm z^2}$ reduces by around 20\%, followed by an increase 
of the PDOS on the C atoms bonded to Mo$_{\rm AB}$  (C$_{\rm AB}$), 
Figs.\,\ref{pdos}(c) and \ref{pdos}(d).

 In Figs.\,\ref{pdos}(e) and 
\ref{pdos}(f) we present the DOS and PDOS of Ru$_{\rm AA}$/tBG-148. Here, the 
double degeneracy of the levels $e$ and $e^\prime$, composed by Ru-4d$_{\rm 
xy}$/4d$_{\rm x^2-y^2}$ and  Ru-4d$_{\rm xz}$/4d$_{\rm yz}$ have 
been  maintained [Fig.\,\ref{pdos}(e)], while the Ru-4d$_{\rm z^2}$ (lying at 
$E_F - 1.9$\,eV) weakly interact with the tBG-148 host. The net magnetic moment 
of 0.3\,$\mu_{\rm B}$ comes from the partial occupation unbalance of Ru-4d$_{\rm 
xz}$ and 4d$_{\rm yz}$ orbitals, as well as from the 2p$_{\rm z}$ orbitals of 
C$_{\rm AA}$ atoms, Fig.\,\ref{pdos}(f). Similarly to the Mo$_{\rm AB}$/tBG, 
the localized character of Ru$_{\rm AA}$-4d$_{\rm z^2}$ orbitals has been missed 
in Ru$_{\rm AB}$/tBG, Fig.\,\ref{pdos}(g). In addition, we find that spin-down 
states, composed by Ru-4d$_{\rm xz}$ and -4d$_{yz}$ orbitals, becomes 
unoccupied increasing the net magnetic moment to 1.8\,$\mu_{\rm B}$. The 
electronic DOS of the carbon atoms bonded to Ru$_{\rm AB}$, C$_{\rm AB}$, 
increases compared with the other C atoms of the graphene layer 
[Fig.\,\ref{pdos}(h)]; contributing to the net magnetic moment of 
Ru$_{\rm AB}$/tBG.

\begin{figure}[thpb]
\includegraphics[width= 8cm]{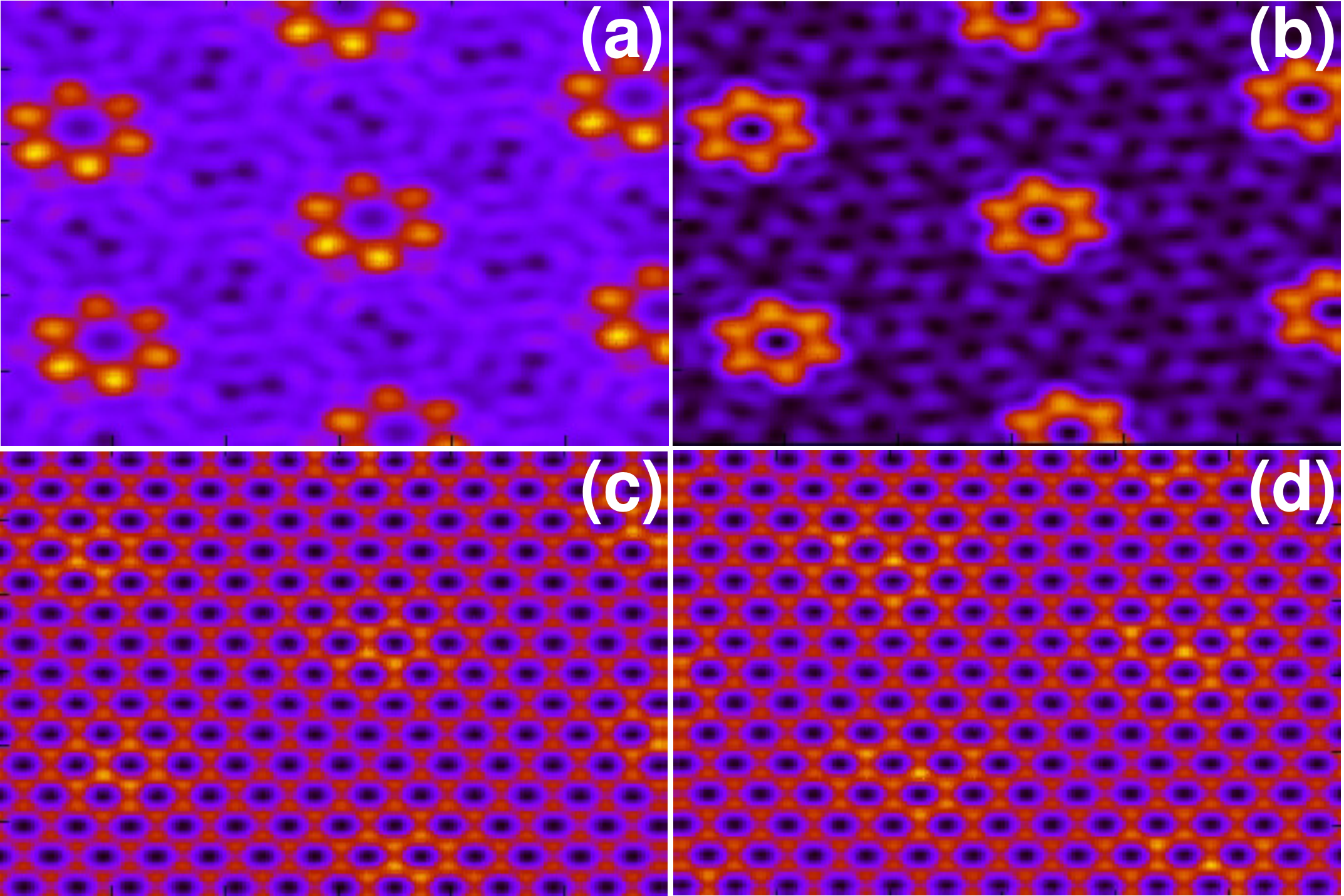}
\caption{\label{stm} Simulated STM images of occupied states ($E_F - 1$\,eV) of a tBG-148 system with 
intercalated Mo$_{\rm AA}$ (a) and Ru$_{\rm AA}$ atoms (b). Simulated STM 
images of the same system without the intercalated atoms but with the corrugation induced by them, 
Mo$_{\rm AA}$ (c) and Ru$_{\rm AA}$ (d).}
\end{figure}

The changes in the electronic density of states on the graphene sheet, induced 
by the intercalated TMs, can be visualized through STM images.  Here, based on 
Tersoff-Hamman\,\cite{tersoff} approach, we simulate a STM image of occupied 
states, within $E_F - 1$\,eV, of Mo$_{\rm AA}$/ and Ru$_{\rm AA}$/tBG-148. As 
depicted in Figs.\,\ref{stm}(a) and \,\ref{stm}(b), we find bright hexagonal spots 
lying on the C atoms above the intercalated Mo$_{\rm AA}$ and Ru$_{\rm AA}$ 
atoms, respectively.  Those spots form a triangular array, and the brightness 
can be assigned to the larger PDOS of C$_{\rm AA}$ atoms  
[Figs.\,\ref{pdos}(b) and \ref{pdos}(f)]. In addition to the electronic 
contribution, the graphene surface corrugation may play an important role. As 
discussed above, the interlayer distance increases slightly upon the 
presence of intercalated TMs, e.g. $\rm\Delta z =  3.69 \rightarrow 3.73$\,\AA\ 
for Mo and Ru in the AA stacking region.

In order to verify the importance of 
such a corrugation on the STM images,  we  simulated STM images of the twisted 
graphene (tBG-148) by removing the intercalated atoms; but keeping the 
equilibrium geometry of the Mo$_{\rm AA}$/ and Ru$_{\rm AA}$/tBG-148 systems. 
Our simulated STM results presented in Figs.\,\ref{stm}(c) and \ref{stm}(d) show 
that indeed  the bright spots follow the   topographic corrugation of the 
graphene  surface.

Several studies have been 
addressed to the formation of magnetic  clusters at the graphene/surface 
interface\,\cite{vo-vanAPL2011,deckerPRB2013, 
deckerJPhysCondMat2014,vlaicAPL2014,limaChemMat2014}. In those systems, the 
formation of (self-assembled) periodic structures is ruled by a suitable synergy 
between the hexagonal lattice of the graphene layer, and the atomic geometry of 
the surface.  Our results suggest that the array of Mo and Ru atoms 
intercalated in the twisted graphene layers can act as a seed to the formation 
of these type of clusters. Indeed, our preliminary results indicate that the 
formation of Ru-dimers intercalated in the AA stacking region is quite likely. 
For Ru-dimers we find $E^b$ of 4.00\,eV/atom,  whereas for Ru-trimers we 
obtained 4.05\,eV/atom. Further studies regarding the formation of TM clusters 
in twisted graphene layers are in progress.

\section{Summary}

We have performed an {\it ab initio}  investigation of TM atoms (Mo, Ru, Co and 
Pt) intercalated in twisted graphene layers.  Our results show that Mo and 
Ru are significantly more stable energetically at tBG AA stacking region in 
comparison with the rest of the unit cell. For Co and Pt atoms the energy differences 
between regions are not relevant. These results indicate that it is possible 
a formation of a triangular  array of Mo and Ru atoms, ruled by the 
twist angle between the graphene layers. In contrast, formation of such a 
structure is not expected for Co and Pt atoms. The energetic 
stability of Mo and Ru atoms at tBG AA stacking regions can be explained in 
terms of the coordination number of the TM atoms with carbon atoms  and also differences in strain 
energy. The net magnetic moments of Mo is 
 quenched upon its intercalation in AA or AB regions, while Ru magnetic moment is 1.9 in the adsorbed system
 and goes to 0.3\,$\mu_{\rm B}$ when intercalated in the 
 AA region. There is an increase of the electronic 
density of states, near the Fermi level, on the graphene carbon atoms bonded to 
the TM atoms, identified through STM images. We find the formation of a 
triangular array of hexagonal bright spots lying on the C atoms nearest neighbour 
to the intercalated Mo and Ru atoms. These spots in the STM images may be used 
 to experimentally identify the intercalated atoms positions.



The authors acknowledge financial support from the Brazilian agencies
CNPq, CAPES and FAPEMIG, the computational support from CENAPAD/SP and 
Chilean FONDECYT grant 1130950.


\section{Appendix}

\begin{figure}[thpb]
\includegraphics[width= 8cm]{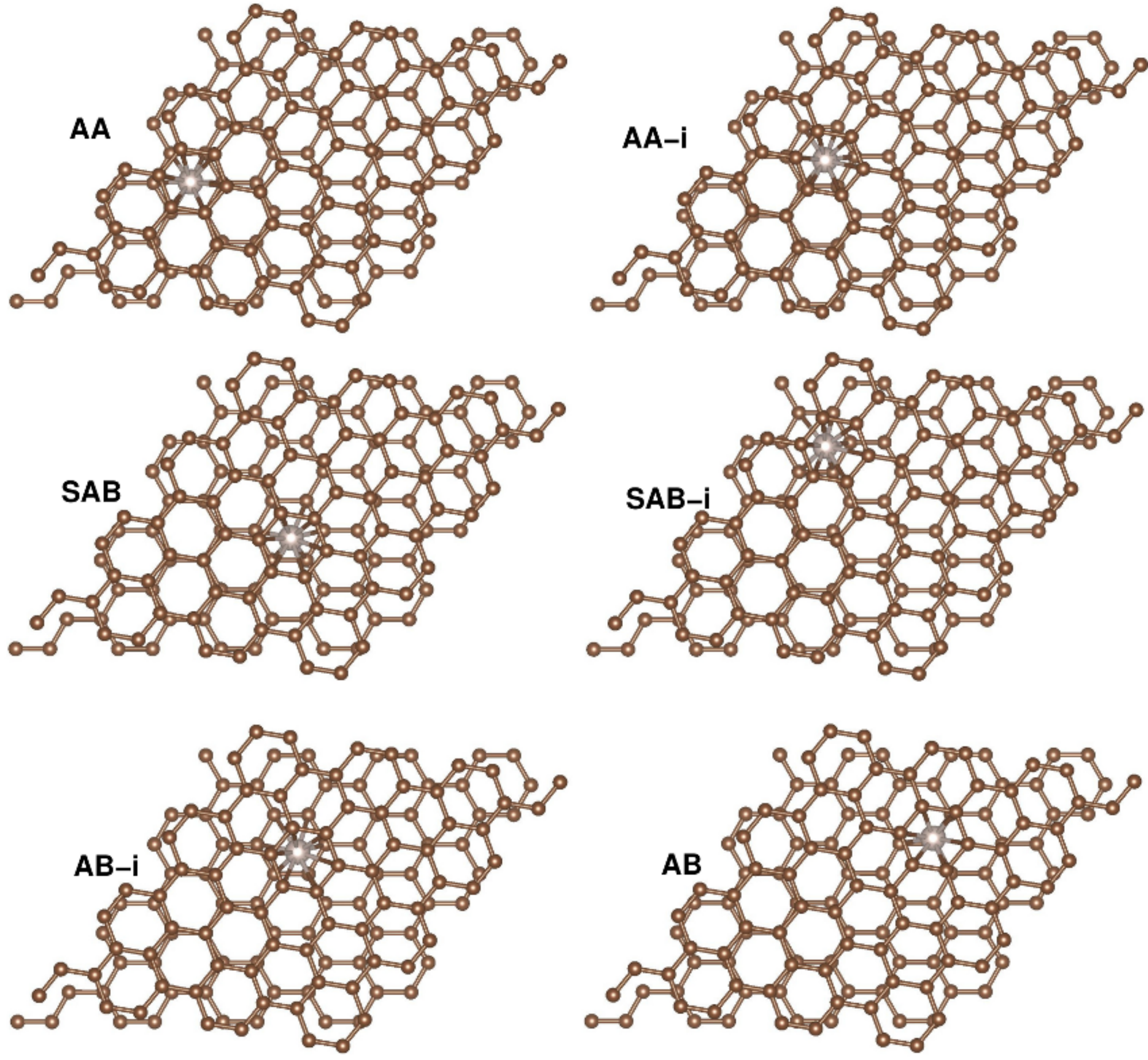}
\caption{\label{Fig5}  Structural models of TM/tGB-148, for TM=Ru and Mo
between the energetically more (less) stable TM$_{\rm AA}$/TGB-148  (TM$_{\rm 
AB}$/TGB-148) configurations.}
\end{figure}

In Fig.\,\ref{Fig5} we show other positions of the intercalated TM atoms
inside the tBG unit cell(tGB-148). The results are for Ru and Mo intercalated atoms since these are
the systems with large energy differences between AA and AB regions. In these cases in order to 
preserve the RRA unchanged during the atomic relaxation process, the C atoms of 
graphene layers are allowed to relax only along the z-direction; such a constrain is not 
applied to the intercalated TMs. In 
Table~\ref{energy-z} we present the total energy differences ($\Delta E$) with 
respect to the energetically more stable TM$_{\rm AA}$ position, 
$$
\Delta E = E[{\rm TM_{AA}}] - E[{\rm TM_{X}}],
$$
where $E[\rm TM_{AA}]$ represents the total energy when the TM is located at the 
AA region, and $E[\rm TM_{X}]$ is the total energy of the other locations (X = 
AA-i, SAB, SAB-i, AB, and AB) depicted in Fig.\,\ref{Fig5}.  These results 
indicate that indeed, there is an energetic preference for the AA sites. The 
total energy difference ($\Delta E$) increases from TM$_{\rm AA}$ to TM$_{\rm 
AB}$. The results of Tables I and III lead us to conclude  that the in plane 
relaxation ($x$ and $y$ direction) plays a minor role on the energetic 
preference of TM atoms. In Table I, where in plane relaxation is considered, the 
results are very close to the ones obtained in Table III where it has not been 
allowed.

\begin{table}[h]
\caption{\label{energy-z}. Total energy difference, $\Delta E = E[{\rm TM_{X}}] 
- E[{\rm TM_{AA}}]$ (in eV) between the energetically more stable TM$_{\rm 
AA}$/TGB-148 (TM = Ru and Mo), and the TM$_{\rm X}$ ones presented in 
Fig.\,\ref{Fig5}.}
\begin{ruledtabular}
\begin{tabular}{lcc} 
 TM$_{\rm X}$      &  Ru  &  Mo  \\
\hline
 TM$_{\rm AA}$     & 0.000 & 0.000 \\
 TM$_{\rm AA-i}$   & 0.085 & 0.120 \\
 TM$_{\rm SAB}$    & 0.168 & 0.249 \\
 TM$_{\rm SAB-i}$  & 0.290 & 0.541 \\
 TM$_{\rm AB-i}$   & 0.296 & 0.478 \\
 TM$_{\rm AB}$     & 0.583 & 1.185 \\
  \end{tabular}
\end{ruledtabular}
\end{table}

\begin{figure}[thpb]
\includegraphics[width= 8cm]{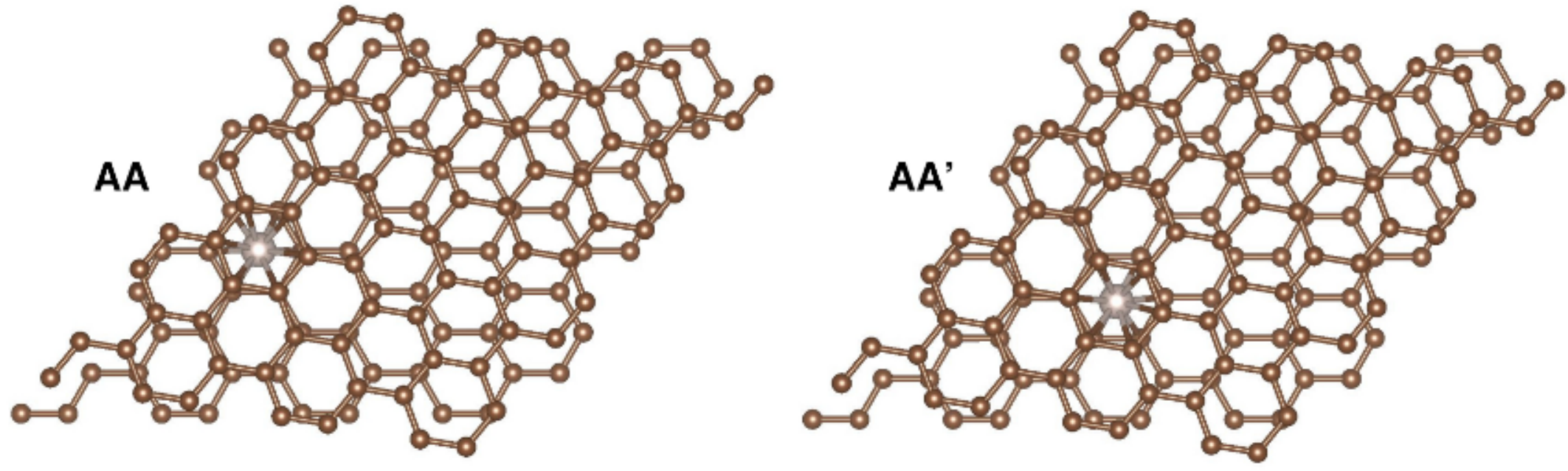}
\caption{\label{Fig6}  Structural models of TM/tGB-148, for TM=Ru and Mo
atoms embedded in AA-like regions, AA and AA$^\prime$.}
\end{figure}

In addition, we compare the total energies for Ru/ and Mo/tGB-148 embedded in  
AA-like regions, indicated as AA and AA$^\prime$ in Fig.\,\ref{Fig6}.
The atomic positions of the TM and tGB are free to relax. Those AA and 
AA$^\prime$ configuration are very close in energy; we find 
differences of 0.007 and 0.024\,eV.



\begin{thebibliography}{44}
\bibitem [{coo()}]{coordination}%
  \BibitemOpen
  \href@noop {} {}\bibinfo {note} {We consider that a TM atom is bonded to a
  carbon atom when the distance between then is smaller or equal to 2.5
  \AA}\BibitemShut {NoStop}%
\end{thebibliography}
%


\end{document}